# Tunable kinetic proofreading in a model with molecular frustration

Andre M. Lindo[1], Bruno F. Faria[2] and Fernao V. de Abreu[2]

[1] ICBAS and I3S, Universidade do Porto, Portugal
[2] Depto de Física and I3N, Universidade de Aveiro, Portugal
{andre.lindo, brunoffaria,fva}@ua.pt

**Abstract.** In complex systems, feedback loops can build intricate emergent phenomena, so that a description of the whole system cannot be easily derived from the properties of the individual parts. Here we propose that inter-molecular frustration mechanisms can provide non trivial feedback loops which can develop nontrivial specificity amplification. We show that this mechanism can be seen as a more general form of a kinetic proofreading mechanism, with an interesting new property, namely the ability to tune the specificity amplification by changing the reactants concentrations. This contrasts with the classical kinetic proofreading mechanism in which specificity is a function of only the reaction rate constants involved in a chemical pathway. These results are also interesting because they show that a wide class of frustration models exists that share the same underlining kinetic proofreading mechanisms, with even richer properties. These models can find applications in different areas such as evolutionary biology, immunology and biochemistry.

## 1 Introduction

An important issue in biochemistry consists in unveiling mechanisms that increase specificity in highly degenerate reactions. A major breakthrough was achieved by Hopfield in 1974 after proposing the kinetic proofreading (KP) mechanism (Hopfield 1974). The KP attempts to explain how DNA replication and protein synthesis occur with very small errors, even though the energies involved in the biochemical recognition processes are very close. The range of possible applications of these ideas was later recognized to be much broader, and opened an active field of research (Hlavacek et al. 2001; Qian 2006; Burroughs and Merwe 2007; Inoue and Kaneko 2010; Owens 2010; Kirkilionis 2010). In 1995, McKeithan argued that the KP could also help explaining the high specificity reached in cellular recognition processes during T cell activation (McKeithan 1995). In this case, highly specific cellular interactions could benefit from the same type of underlying mechanism which helped explaining the high specificity achieved in molecular interactions (Chan et al. 2003; George 2005). In a quite different context one of the authors of this contribution identified a mechanism that highly increased assortativeness in a mating population. Assortativeness is another kind of specific interaction between individuals which favors some matings relatively to others. In (Almeida and de Abreu 2003) it was argued that these specifically amplified interactions could explain the emergence of new species in sympatry. Later, it was also proposed that similar ideas could be applied in immunology (de Abreu et al. 2006) in what it was called the cellular frustration framework. However, these later models are substantially different from Hopfield's and McKeithan's KP models. In the later case increased specificity depends only on the specific rate constants associated to each molecular interaction, whereas in (de Abreu et al. 2006) interactions depend non-trivially on the presence of other agents (molecules, cells or individuals).

Since both types of models can appear to be very different, it is important to show that they both share the same underlying mechanism, namely, the existence of a long series of time lasting interactions until a final reaction stage is reached. The purpose of this paper is thus to show that the mechanism identified in (Almeida and de Abreu 2003) can be seen as a general form of a kinetic proofreading mechanism. For this purpose we organize this paper as follows. In the next section we will discuss the main ideas involved in the kinetic proofreading mechanism as discussed by McKeithan. McKeithan's model is extremely appealing due to its simple mathematical formulation. We will concentrate on those aspects that allow a straightforward comparison between both classes of models. Afterwards, we introduce a simple model that captures the main ideas of cellular frustration models and discuss a molecular version of these models. We will show that it is possible to use the formulation introduced by McKeithan within this more general class of models. We will be able to show that in the later models increased specificity can indeed be achieved, and furthermore, that this class of models can exhibit a new property. This is the possibility of tuning specificity amplification by changing reactants concentrations. This result is extremely interesting because, the classical kinetic proofreading pathway did not benefit from this type of flexible behavior.

## 2   Classical Kinetic Proofreading

Consider three types of molecules A, $A^*$ and B. Molecules A and $A^*$ can both react with B molecules with very close free energies. The KP mechanism shows that it is possible to produce an almost unlimited larger number of final products from one reaction than the other. The KP mechanism makes two important assumptions. It assumes that the final products are obtained after a long chain of reactions, and also that intermediate reactants can be destabilized towards the initial step in the pathway, as described in Fig. 1. If destabilization rates are bigger for reactions involving one molecule (A or A$^*$) than the other, then a several fold increase in the production of the final products can be achieved in one pathway relatively to the other. This is the essence of KP explanation for the emergence of high specificity arising from almost degenerate reactions.

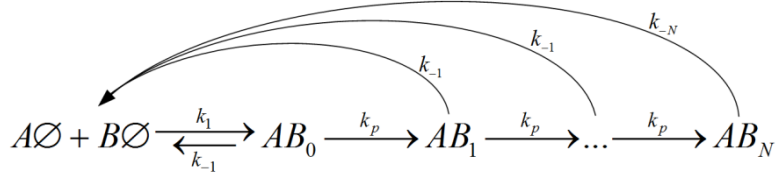

**Fig. 1.** Classical kinetic proofreading pathway as described by McKeithan. Initially free molecules A and B (respectively represented by AØ and BØ), react forming intermediate complexes $AB_i$. An identical pathway exists to produce $A^*B_N$ final products, but reaction rate constants differ. In particular, the KP mechanism requires that a difference exists in dissociation constants $k_{-1}$ and $k_{-1}^*$. The association rate constant is $k_1$.

From a mathematical point of view the specificity increase can be straightforwardly understood. Following McKeithan (1995), we would write the chemical kinetics equations:

$$\frac{d[AB_0]}{dt} = k_1 \cdot [A\emptyset][B\emptyset] - k_{-1} \cdot [AB_0] - k_p \cdot [AB_0] \ . \tag{1}$$

$$\frac{d[AB_i]}{dt} = k_p \cdot [AB_{i-1}] - k_{-1} \cdot [AB_i] - k_p \cdot [AB_i] \ . \tag{2}$$

$$\frac{d[AB_N]}{dt} = k_p \cdot [AB_{N-1}] - k_{-1} \cdot [AB_N] \ . \tag{3}$$

and similarly for the pathway involving the A$^*$ molecule. The total number of molecules A and B is assumed constant. As a result the concentration of free molecules evolves according to $d[A\emptyset]/dt = d[B\emptyset]/dt = -\Sigma d[AB_i]/dt$ ($i=0,\ldots,N$).

It is easy to obtain the following relations for the concentrations at the steady-state:

$$\frac{[AB_i]_{ss}}{[AB_{i-1}]_{ss}} = \frac{k_p}{k_p + k_{-1}} \equiv \alpha \ . \tag{4}$$

$$\frac{[AB_N]_{ss}}{[AB_0]_{ss}} = \alpha^{N-1} \cdot \frac{k_p}{k_{-1}} \ . \tag{5}$$

$$\frac{[AB_i]_{ss}}{[AB_0]_{ss}} = \alpha^i .\tag{6}$$

The total concentration of complexes ($[AB_{total}]$) can be obtained by adding (5) and the summation of all the other complexes concentrations using (6):

$$[AB_{total}]_{ss} = [AB_0]_{ss} \cdot \left( \alpha^{N-1} \cdot \frac{k_p}{k_{-1}} + \sum_{i=0}^{N-1} \alpha^i \right) = [AB_0]_{ss} \left( 1 + \frac{k_p}{k_{-1}} \right) .\tag{7}$$

Finally, it is possible to write for the normalized concentration of final products:

$$\frac{[AB_N]_{ss}}{[AB_{total}]_{ss}} = \alpha^N .\tag{8}$$

Identical expressions can be obtained for the pathway involving $A^*$ molecules, replacing, $k_{-1}$ by $k_{-1}^*$, or, correspondingly, $\alpha$ by $\alpha_*$.

To quantify the effectiveness of the kinetic proofreading mechanism, we follow (Chan et al. 2003) and define true positive (TP) events when $A^*B_N$ complexes are formed. On the other hand, false positive (FP) events occur when $AB_N$ complexes are formed. True negative (TN) and false negative (FN) events occur every time $AB_0$ or $A^*B_0$ complexes do not end in $AB_N$ or $A^*B_N$ complexes, respectively. The following statistical measures of performance can then be defined:

$$\text{Specificity} = \frac{\text{Number of TP}}{\text{Number of TP} + \text{Number of FP}} .\tag{9}$$

$$\text{Sensitivity} = \frac{\text{Number of TP}}{\text{Number of TP} + \text{Number of FN}} .\tag{10}$$

$$\text{Tolerance} = \frac{\text{Number of TN}}{\text{Number of TN} + \text{Number of FP}} .\tag{11}$$

Specificity measures the fraction of final products, $AB_N$, that corresponds to the desired complex. Sensitivity measures the fraction of complexes $A^*B_0$ that produce the desired final product, $A^*B_N$. Finally, tolerance measures the fraction of complexes $AB_0$ that do not form the undesired final product, $AB_N$.

Tolerance was not considered in (Chan et al. 2003) because in McKeithan's model tolerance can be expressed as a function of sensitivity and specificity. However, we include it here because its behavior will be useful to discuss separate phenomena occurring in the molecular frustration model. From (7) and (8) we have:

$$\text{Sensitivity} = \left( \frac{k_p}{k_p + k_{-1}^*} \right)^N = \alpha_*^N .\tag{12}$$

$$\text{Tolerance} = 1 - \left(\frac{k_p}{k_p + k_{-1}}\right)^N = 1 - \alpha^N. \tag{13}$$

$$\text{Specificity} = \frac{\alpha_*^N}{\alpha_*^N + \alpha^N}. \tag{14}$$

Since $\alpha < \alpha_* < 1$, it is clear that sensitivity decreases exponentially with the number of reaction steps $N$, while specificity approaches 1. Specificity amplification depends only on the dissociation rate constants, $k_{-1}$ and $k_{-1}^*$, their ratio and absolute values (Table 1). It should be remarked that according to McKeithan's scheme, specificity is not necessarily close to 1. For small values of $k_{-1}/k_p$ specificity can be only slightly higher than 50%. To increase specificity it is required to increase dissociation rate constants. However, this has a drawback, as it reduces the overall sensitivity (reactivity) of the system (Table 1).

According to McKeithan's scheme, specificity depends only on chemical constants and cannot be easily changed. In the next section, we will show that molecular frustration also increases specificity, but in that case specificity depends on the reactants concentrations, and hence it can be easily tuned.

**Table 1.** Specificity for several systems with 4 states ($N=3$) in the pathway.

| $k_{-1}^*/k_p$ | $k_{-1}/k_P = 100$ | | $k_{-1}/k_P = 1$ | |
|---|---|---|---|---|
| | 1 | 10 | 0.01 | 0.1 |
| Tolerance | $\cong 1.0000$ | $\cong 1.0000$ | 0.8750 | 0.8750 |
| Specificity | $\cong 1.0000$ | 0.9987 | 0.8859 | 0.8574 |
| Sensitivity | 0.1250 | 0.0008 | 0.9706 | 0.7513 |

## 3 Inter-Molecular Frustration

An alternative mechanism to increase specificity in systems of complex agents was proposed in (de Abreu et al. 2006), in the context of a theory for sympatric speciation. Instead of molecules involved in multi-step chained reactions, it was considered a population of individuals performing time lasting mating decisions. This approach shares a common ingredient with the classical KP mechanism, namely the requirement of time demanding intermediate processes preceding the final stage. During these intermediate steps dissociations occur with different probabilities depending on the already formed associations ($A^*B$ or $AB$ in the example from the previous section). However now, dissociations arise depending on which molecules are interacting. As a result dissociation rates become frequency dependent: they depend on the concentrations of molecules that can destabilize a given complex.

According to the classical KP approach, intermediate steps could be due to a set of conformational configurations (McKeithan 1995; de Abreu et al. 2006), or other intermediate chemical reactions (Hopfield 1974). The final step in the pathway would be associated with the production of major signals (like cellular activation (McKeithan 1995; de Abreu et al. 2006), or the assembly of the final synthesis product (Hopfield 1974). On the contrary, in the case of individuals in a mating population, the time consuming intermediate states would arise from the time investment required before a decision is made. The major signal produced in the end could be related to reproduction (Almeida and de Abreu 2003).

However, a fundamental difference exists between the two approaches. In the traditional KP mechanism intermediate steps are destabilized by spontaneous dissociation. This is fixed according to the chemistry of the reactions involved. On the contrary, in the approach taken in (Almeida and de Abreu 2003), this would be due to the other individuals decisions that could change a mating stability. This led to the concept of "cellular frustration" (de Abreu et al. 2006): an individual's decision may frustrate other individual's previously taken decisions. In this work we propose that similar ideas may also be applicable to describe interactions between large molecules or complexes. Furthermore we want to show that the frustration mechanism proposed fits in a more general form of a kinetic proofreading mechanism, which displays new features.

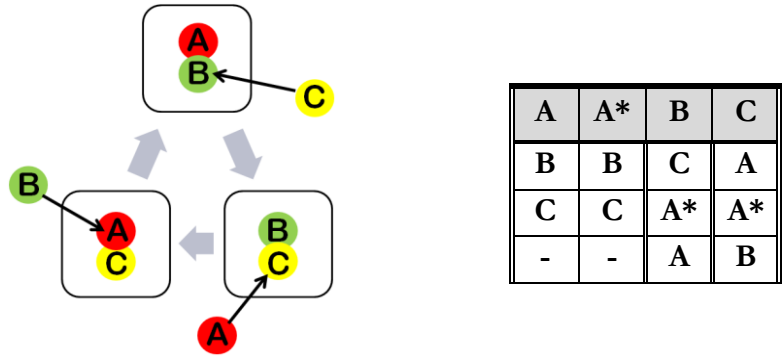

**Fig. 2.** Frustrated dynamics (*left*) displayed by a system of three molecules interacting according to the interaction rules displayed in the table on the *right*. In each column molecules interacting with the molecule in the first line, are ranked in a decreasing affinity order. Accordingly, molecule A binds with higher affinity to molecule B than to molecule C. As a result, if molecule A binded molecule C, and interacts with molecule B (configuration on the left), then it will establish a stable configuration with molecule B and unbind molecule C (top left configuration). In the table we also included a fourth less frustrated molecule $A^*$ which is not shown on the *left*. The set of four molecules forms the model we study in this work.

The main ideas involved in the "cellular frustration" concept can be easily understood with the example in Figure 2. Consider 3 molecules (A, B and C) interacting according to a ranking of affinities as described in the Table in Figure 2 (*right*). For instance, molecules A would bind to molecules B with higher binding energy than if they would bind to molecules C. Similarly, molecules B would bind to molecules C with higher binding energy than to molecules $A^*$ and A, respectively and in decreasing order.

Note that the frustration mechanism discussed here is different from the one discussed in the context of protein folding (Bryngelson et al. 1995). In protein folding, frustration occurs when amino-acid sequences produce many conformations that are closely linked local minima in the energy landscape. Frustration is mainly an intra-molecular concept. In our case, frustration emerges from inter-molecular interactions, producing also many unstable configurations. To distinguish the two types of frustration, we refer to the frustration mechanisms discussed in this paper as inter-molecular frustration.

Crucially important is that it is also assumed that a molecule can only form a stable binding with one other molecule at a time. Consequently, when two molecules are simultaneously bound to a third, it is assumed that this creates an instability reducing drastically the affinity in the weaker bond. This could arise, for instance, if conformations develop whenever stronger bindings are formed, weakening previous ones. While in the case of mating decisions this kind of decision dynamics seems to be most natural, in the context of molecular interactions it requires a bigger complexity in the dynamics of individual molecules. Nevertheless, this is still a reasonable possibility for many systems. And indeed, if one thinks of transmembrane proteins, it is well known that binding extracellular molecules can induce release of intracellular molecules in a far away binding site at a later time (Hlavacek 2001; Burroughs 2007).

Molecular frustration could then be a natural outcome in this kind of complex systems. As shown in Figure 2, molecules A, B and C would tend to form continuously transient bindings because a third molecule can always destabilize already established pairs. The question we raised in this work is whether the introduction of a fourth molecule, different and yet similar to molecule A, may lead to longer lasting (less frustrated) interactions, so that the final interaction step in the chemical pathway is more easily reached?

## 4 Inter-Molecular Frustration and increased specificity

We considered a biochemical model with 4 molecules (A, A$^*$, B and C) interacting according to the Table in Figure 2 (*right*) and undergoing a sequence of conformational steps before the final AB or A$^*$B complexes are reached. We will assume that molecule A is maximally frustrated (de Abreu and Mostardinha 2009) so that the introduction of a new molecule A$^*$ necessarily reduces frustration leading to more stable A$^*$B complexes. This model allows a direct comparison with McKeithan's model, and highlights how molecular frustration embodies a generalization of a kinetic proofreading mechanism.

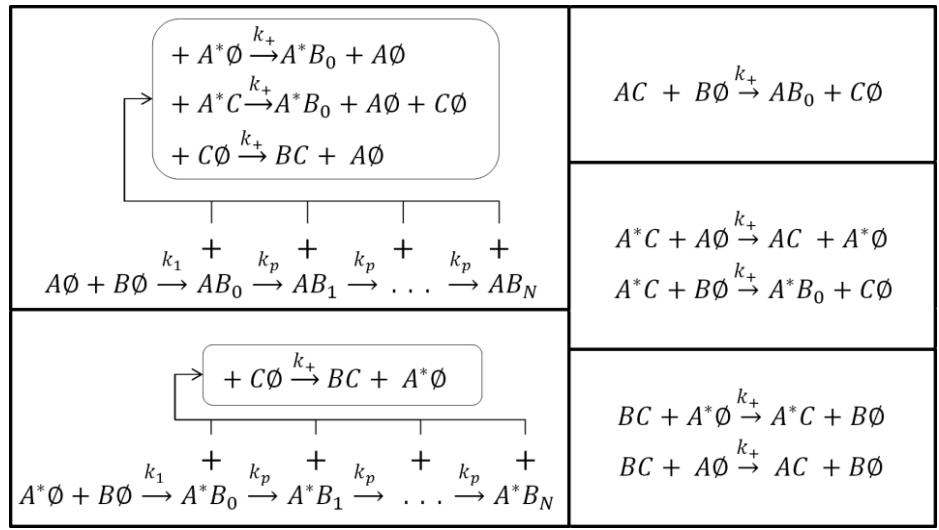

**Fig. 3.** Full set of reactions considered in the model. In the first pathway (top left) AB$_i$ complexes can be dissociated as a result of interactions with free A$^*$ and C molecules (represented respectively by A$^*\varnothing$ and C$\varnothing$), or with A$^*$C complexes. On the contrary, A$^*$B$_i$ complexes can only be destabilized by interactions with free C molecules (bottom left).

In Figure 3 the full set of reactions is presented. If alone, A and B molecules can react with a reaction rate constant $k_+$, forming an $AB_0$ complex. Afterwards they can undergo a sequence of transformations ($AB_i \rightarrow AB_{i+1}$), at a reaction rate constant $k_p$. During these intermediate configurations, AB complexes can also be destabilized, if free C molecules interact with the B molecule in the complex. In this case a new BC complex is formed and an A molecule is freed. Afterwards, the BC complex can be destabilized if a molecule A interacts with the C molecule in the complex. It should be noted that, according to the table in Figure 2, an $A^*$ molecule can also destabilize an AB complex. AB complexes are more frustrated than $A^*B$ and consequently it should be harder that AB complexes reach the final product stage. It is our aim to quantify the specificity increase that is introduced by the fact that $A^*$ molecules are necessarily less frustrated than A molecules.

Note that for the sake of mathematical simplicity we assumed that the $k_+$ reaction constant incorporates two processes. One concerning the binding of two molecules, and the other the release of formerly bonded molecules on a distant binding site. The later could result from a change in the conformational state of that binding site. This simplification allows a more transparent mathematical description and should not influence qualitatively the results provided $k_+ \gg k_p$, i.e., the time scale involved in binding and change on the conformational structure on distant binding sites is a fast process relatively to the time required to reach final products.

Simple mean field dynamical equations can be derived for the concentrations of the several possible complexes. Equations for the $A^*B$ complexes are:

$$\frac{d[A^*B_0]}{dt} = k_+ \cdot \left(\left(\sum_{i=0}^{N}[AB_i]+[B\emptyset]\right) \cdot \left([A^*\emptyset]+[A^*C]\right)\right) \\ - k_p \cdot [A^*B_0] - k_+ \cdot [A^*B_0] \cdot [C\emptyset] \qquad (15)$$

$$\frac{d[A^*B_i]}{dt} = k_p \cdot [A^*B_{i-1}] - [A^*B_i] \cdot \left(k_p + k_+ \cdot [C\emptyset]\right). \qquad (16)$$

$$\frac{d[A^*B_N]}{dt} = k_p \cdot [A^*B_{N-1}] - k_+ \cdot [A^*B_N] \cdot [C\emptyset]. \qquad (17)$$

For equations involving the AB complexes we got:

$$\frac{d[AB_0]}{dt} = k_+ \cdot [B\emptyset] \cdot \left([A\emptyset]+[AC]\right) - k_p \cdot [AB_0] \\ - k_+ \cdot [AB_0] \cdot \left([C\emptyset]+[A^*C]+[A^*\emptyset]\right) \qquad (18)$$

$$\frac{d[AB_i]}{dt} = k_p \cdot [AB_{i-1}] - [AB_i] \cdot \left(k_p + k_+ \cdot [C\emptyset] + [A^*C] + [A^*\emptyset]\right). \quad (19)$$

$$\frac{d[AB_N]}{dt} = k_p \cdot [AB_{N-1}] - k_+ \cdot [AB_N] \cdot \left([C\emptyset] + [A^*C] + [A^*\emptyset]\right). \quad (20)$$

Similar equations were derived for the other complexes (e.g., AC, BC, etc) to simulate the whole system. The concentration of free A molecules is obtained using $d[A\emptyset]/dt = -\Sigma d[AB_i]/dt - d[AC]/dt$ ($i=0,\ldots,N$) and similarly for the other free molecules.

In equations (15-20) negative contributions involving the rate constant $k_+$ account for dissociations resulting from interactions with other molecules. Contrary to what happened in McKeithan's model, the rate constant $k_+$ is now modulated by the concentration of molecules that can destabilize a complex. For instance, the $A^*B_N$ complex can be dissociated upon interaction with a free C molecule (equation 17), whereas the $AB_N$ complex can furthermore be destabilized by $A^*C$ complexes and free $A^*$ molecules. $AB_N$ complexes are thus less stable than $A^*B_N$ complexes. By performing a similar analysis to equations (15-16) and (18-19) we can conclude in general that AB complexes are more unstable than A*B complexes as there are more processes destabilizing the former complexes (Figure 3).

At the steady-state, $d[A^*B_i]/dt = d[AB_i]/dt = 0$, and we can define constants $\beta$ similar to McKeithan's $\alpha$ constants:

$$\frac{[A^*B_i]_{ss}}{[A^*B_{i-1}]_{ss}} = \frac{k_p}{k_p + k_+ \cdot [C\emptyset]_{ss}} = \beta_* . \quad (21)$$

$$\frac{[AB_i]_{ss}}{[AB_{i-1}]_{ss}} = \frac{k_p}{k_p + k_+ \cdot \left([C\emptyset]_{ss} + [A^*C]_{ss} + [A^*\emptyset]_{ss}\right)} = \beta . \quad (22)$$

A relation can also be obtained relating the concentrations at the initial and the last step in the pathway:

$$\frac{[A^*B_N]_{ss}}{[A^*B_0]_{ss}} = \frac{k_p}{k_+ \cdot [C\emptyset]_{ss}} \cdot \beta_*^{N-1} . \quad (23)$$

Finally, the total concentration of $A^*B$ complexes in the system is given by:

$$[A^*B_{total}] = [A^*B_0]_{ss} \left( \beta_*^{N-1} \cdot \frac{k_p}{k_+ \cdot [C\emptyset]_{ss}} + \sum_{i=0}^{N-1} \beta_*^i \right) = [A^*B_0]_{ss} \left( 1 + \frac{k_p}{k_+ \cdot [C\emptyset]_{ss}} \right). \quad (24)$$

Expressions are identical to the ones obtained by McKeithan if we establish the following correspondences:

$$k^*_{-1} \leftrightarrow k_+ \cdot [C\emptyset]_{ss} \ , \qquad k_{-1} \leftrightarrow k_+ \cdot \left([C\emptyset]_{ss} + [A^*C]_{ss} + [A^*\emptyset]_{ss}\right). \qquad (25)$$

Similarly, expressions for the sensitivity, specificity and tolerance remain identical if we replace $\alpha \to \beta$ and $\alpha_* \to \beta_*$. An important consequence of this analysis is that molecular frustration can also produce an increase in specificity similar to that predicted by the classical KP mechanism. However, there is an important difference: specificity is now a function of reactants steady state concentrations. This is interesting for two main reasons. First, because it shows that the mechanism of inter-molecular frustration embodies a more general kind of a kinetic proofreading mechanism that does not require a numerous sequence of conformational pathways. Instead it builds upon the emergent behavior of a fairly simple set of inter-molecular interactions. Consequently, this calls attention to the fact that self-organized molecular interactions (Karsenti 2008) can achieve specificity amplification similar to what happened in the less flexible KP schemes. Secondly, because it shows that there are mechanisms in which specificity could be tuned to a desired level by an external control, or by a mechanism developed inside a complex biochemical system, for instance a regulated synthesis of C molecules. Hence, molecular frustration can be a flexible specificity amplification mechanism.

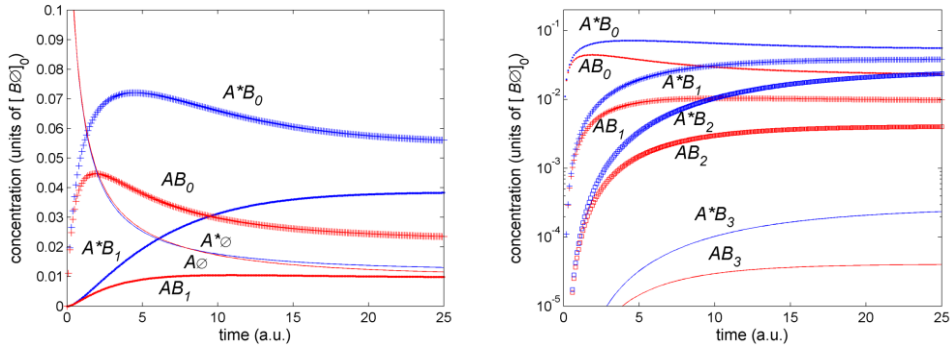

**Fig. 4.** The evolution of the concentrations for unbounded molecules A and $A^*$ ($A\emptyset$ and $A^*\emptyset$) and intermediate complexes $AB_i$, $A^*B_i$ ($i=0,...,N$) in a system with $N=3$, $k_+/k_p=10$. In the initial configuration there were only unbound molecules with relative concentrations: $[A\emptyset]_0/[B\emptyset]_0=[A^*\emptyset]_0/[B\emptyset]_0=[C\emptyset]_0/[B\emptyset]_0=0.5$.

In order to confirm these theoretical predictions, we simulated numerically the reaction rate dynamical equations derived before. In Figure 4 we show the evolution of the concentration of $AB_i$ and $A^*B_i$ complexes for $i=0,...,N$, together with $[A\emptyset]$ and $[A^*\emptyset]$. It is clear that the concentration of single molecules decreases straightforwardly along the time while the concentration at the steady state of intermediate complexes decreases steadily along the pathway. As a result, while there is a specificity increase, there is simultaneously an overall loss in reactivity, typical of kinetic proofreading mechanisms (Chan et al. 2003; George 2005).

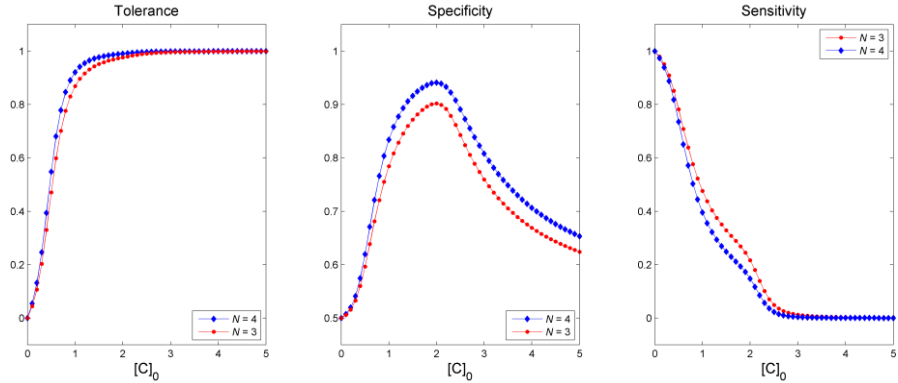

**Fig. 5.** Dependence of tolerance *(left)*, specificity (*center*) and sensitivity (*right*) when the initial concentration of a third molecule C is varied. Considered systems were simulated for $k_+/k_p=10$ with $[A\emptyset]_0/[B\emptyset]_0=[A^*\emptyset]_0/[B\emptyset]_0=0.5$ with $N=3$ and $N=4$.

On Figure 5 we show the dependence of tolerance, specificity and sensitivity as a function of the initial concentration of C molecules and for pathways with $N=3$ and $N=4$. These results are interesting for several reasons. First because they confirm that specificity varies considerably with the concentration of one of the reactants. In particular, it shows that the introduction of a third molecule influences the specificity dramatically. As a result, specificity becomes a tunable property because it can be increased to a desired level by tuning the third molecule concentration, at the expense of some decrease in sensitivity. These results are also interesting because specificity shows a maximum at a location that does not coincide with the maximal increase in tolerance. Furthermore, sensitivity is still non-negligible when tolerance approaches the maximal value. This puts in evidence that there are two different mechanisms taking place. Specificity amplification uses the same mechanism leading to increased tolerance. Indeed, both require frustration to build the necessary feedback loops. However, if the concentration of C reactants is too large, a jamming transition takes place that again blocks these amplification mechanisms, reducing both sensitivity and specificity. The non-monotonous decay in sensitivity plot in Figure 5 indeed shows that these two mechanisms are present.

In Figure 6 we also show that specificity increases when the molecular reactivity, $k_+$, increases. This agrees with (25) and highlights that specificity is an emergent property in this model. In fact, one could naively think that by increasing the molecular reactivity of all species, specificity could be hindered. On the contrary specificity increases and the production of some species over others can differ by orders of magnitude. In this figure the amplification ratio was computed according to:

$$\text{Amplification} = \frac{[A^*B_N]}{[A^*\emptyset]} \bigg/ \frac{[AB_N]}{[A\emptyset]}. \tag{26}$$

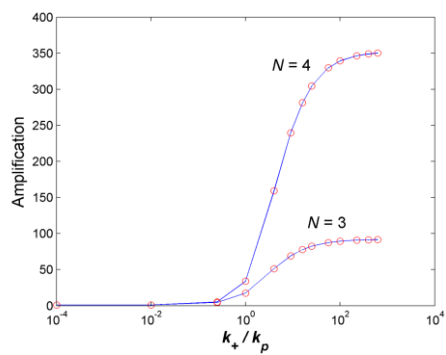

**Fig. 6.** The amplification in specificity defined in (29), as a function of the molecular reactivity $k_+$ for systems with $[C\emptyset]_0/[B\emptyset]_0=2$, $[A\emptyset]_0/[B\emptyset]_0=[A^*\emptyset]_0/[B\emptyset]_0=0.5$ and $N=3$ and $4$.

## 4 Final Discussion and Conclusions

In this article we discussed a possible mechanism for specificity amplification based on molecular frustration. Molecular frustration requires that each molecule has at least two different binding sites for different molecules. It also requires that each molecule can only form stable bindings with one single molecule at a time. If it binds to a third molecule, then the weaker bond becomes unstable. This could result from a kind of 'induced fit' mechanism, in which after the initial binding conformations follow and destabilize other binding sites, a process resembling long-ranged allosteric regulation (Kenakin et al. 2010; Gandhi et al. 2008). It is, however, not possible to exclude a 'conformation selection' scenario, probably of an extended type (Csermely et al. 2010), in which both conformational selection and adjustments follow each other. In fact, both mechanisms are likely to coexist in many cases (Hammes et al. 2009). In any way, we believe that the specificity amplification phenomenon here discussed, is robust against the particular type of binding events, 'induced fit' or 'conformational selection'.

Inter-molecular frustration arises if a sequence of bindings and unbindings occurs so that molecules never form stable complexes. In case the set of molecules in the system forms a maximally frustrated set (de Abreu and Mostardinha 2009), we showed that a less frustrated molecule introduced in the system can produce final products at a rate one or two fold higher than products involving other similar molecules in the system. Furthermore we showed that this specificity amplification depended on the reactants concentrations. This result is interesting because it contrasts with the conventional kinetic proofreading mechanism in which specificity is a sole function of chemical rate constants and consequently it is not as flexible. It would be interesting to consider other biochemical systems displaying similar increased specificity properties, but with different biochemical requirements.

Finally, the results reported in this article also established a clear link between kinetic proofreading mechanisms, well known in biochemistry or theoretical immunology, and assortativeness amplification mechanisms relevant in evolutionary biology (Almeida and de Abreu 2003).